\documentclass[11pt]{article}
\pdfoutput=1
\usepackage{amsmath,amssymb,color,epsfig,cite}
\usepackage{graphicx}
\usepackage{subfigure}
\usepackage{setspace}

\usepackage{amsfonts}

\newcommand{\hoch}[1]{$\, ^{#1}$}

\makeatletter
\@addtoreset{equation}{section}
\makeatother

\newcommand{\be}{\begin{equation}}
\newcommand{\ee}{\end{equation}}
\newcommand{\bea}{\setlength\arraycolsep{2pt} \begin{eqnarray}}
\newcommand{\eea}{\end{eqnarray}}

\def\0{{\sst{(0)}}}
\def\1{{\sst{(1)}}}
\def\2{{\sst{(2)}}}
\def\3{{\sst{(3)}}}
\def\4{{\sst{(4)}}}
\def\5{{\sst{(5)}}}
\def\6{{\sst{(6)}}}
\def\7{{\sst{(7)}}}
\def\8{{\sst{(8)}}}
\def\sst#1{{\scriptscriptstyle #1}}

\begin{document}

\begin{center}
{\large {\bf ${\cal P}-v$ Criticality in Gauged Supergravities}}

\vspace{15pt}
{\large Shou-Long Li\hoch{1,2}, Hong-Da Lyu\hoch{3},
Hua-Kai Deng\hoch{1} and Hao Wei\hoch{1}}

\vspace{15pt}

\hoch{1}{\it School of Physics, Beijing Institute of Technology, Beijing 100081, China }

\vspace{10pt}

\hoch{2}{\it Center for Joint Quantum Studies and Department of Physics,\\
School of Science, Tianjin University, Tianjin 300350, China }

\vspace{10pt}

\hoch{3}{\it Department of Physics, Beijing Normal University, Beijing 100875, China}

\vspace{40pt}

\underline{ABSTRACT}
\end{center}

AdS black holes show  richer transition behaviors in extended phase space  by assuming the cosmological constant and its conjugate quantity to behave like thermodynamic pressure and thermodynamic volume. We study the extended thermodynamics of charged dilatonic AdS black holes in a class of Einstein-Maxwell-dilaton theories that can be embedded in  gauged supergravities in various dimensions. We find that the transition behaviors of higher dimensional dilatonic AdS black holes are different from the four dimensional counterparts, and new transition behaviors emerges in higher dimensions. First, there exists standard Van der Waals transition only in a five dimensional dilatonic AdS black hole with two equal charges. Second, there emerge a new phase transition branch in negative pressure region in six and seven dimensional dilatonic black holes with two equal charges. Third, there emerge transition behaviors in higher dimensional black hole with single charge cases, which are absent in four dimensions.

\vfill
 sllee\_phys@bit.edu.cn\ \ \ hongda\_lv@163.com \ \ \  dhklook@163.com \ \ \ haowei@bit.edu.cn

\thispagestyle{empty}

\pagebreak



\newpage

\section{Introduction}

In view of the anti-de Sitter/conformal field theory (AdS/CFT) correspondence~\cite{Maldacena:1997re, Gubser:1998bc, Witten:1998qj}, the strongly coupled boundary conformal field theory (CFT) can be understood via anti-de Sitter (AdS) black holes. In terms of the AdS/CFT correspondence, the Hawking-Page phase transition~\cite{Hawking:1982dh} of Schwarzchild AdS black hole was well interpreted by Witten~\cite{Witten:1998zw} as a confinement/deconfinement phase transition in the boundary CFT. However, the dual CFT of the phase transition, the Van der Waals transition, of a Reissner-Nordstr\"om (RN) AdS black hole is still unclear. So it is valuable to study the phase transitions thoroughly in  both boundary and bulk points of view. The phase transitions of charged AdS black hole systems have been studied in the literature, e.g.~\cite{Chamblin:1999tk, Chamblin:1999hg,Cvetic:1999ne,Cvetic:1999rb }, for many years. Recently, the idea of treating the cosmological constant and its conjugate quantity on the same footing as thermodynamic pressure and thermodynamic volume in first law of black hole thermodynamics led to studies in Refs.~\cite{Caldarelli:1999xj,Kastor:2009wy, Cvetic:2010jb} and next the phase transition in extended phase space was considered in Ref.~\cite{Kubiznak:2012wp}. In the frame of extended phase space, many interesting new features appear, for example,  $\lambda$-line transition~\cite{Hennigar:2016xwd}, reentrant phase transitions~\cite{Gunasekaran:2012dq,Altamirano:2013ane}, triple points~\cite{Altamirano:2013uqa}, special isolated critical point~\cite{Dolan:2014vba} and so on~\cite{Hennigar:2015esa, Hendi:2017fxp}. We refer to e.g.~\cite{Kubiznak:2016qmn} and the references therein for more details of this subject.

In the transitions mentioned above, the effects of the scalar fields are not considered. And one of the simplest scalar modifications of Einstein-Maxwell (EM) theory is the so-called Einstein-Maxwell-dilaton (EMD) theory. There are many different EMD theories due to different dilaton coupling constants and scalar potentials, and phase transitions are studied in these theories, e.g. in \cite{Hristov:2013sya, Zhao:2013oza, Dehghani:2014caa}. In our work, we focus on a class of EMD theory inspired by supergravity.

The extremal RN black holes in supergravity can be viewed as one of the bound states of the basic $U(1)$ building blocks with zero binding energy~\cite{Duff:1994jr, Rahmfeld:1995fm}. On the other hand, while EM theories can be embedded in string and M-theory in four and five dimensions only, charged dilatonic AdS black holes in gauged supergravities~\cite{Behrndt:1998jd, Duff:1999gh, Cvetic:1999xp, Cvetic:1999un} can be embedded in higher dimensions.

Actually, according to Ref.~\cite{Caceres:2015vsa}, there exist different phase transitions for charged dilatonic AdS black holes with different dilaton coupling constants in EMD theory with a string-inspired potential in four dimensions. The dilaton coupling constants can be reparametrized by the parameter $N$ (which is given by Eq.~(\ref{constraint1})), and the positive integer value of $N$ related to the number of equal charges of the dilatonic AdS black holes. So it is natural to ask whether the higher dimensional dilatonic AdS black holes with $N$ equal charges have the same transition behaviors  as their four dimensional counterparts, and whether  there is any new transition behavior emerging in higher dimensions.   We will explore the answers in this paper.

The rest of this paper is organized as follows. In section~\ref{sec2}, we review the thermodynamics of charged dilatonic AdS black holes in gauged supergravity. In section~\ref{sec3}, we study the phase transitions  of charged dilatonic AdS black holes in extended phase space of canonical ensemble in various dimensions and dilaton coupling constants. We conclude in section~\ref{sec4}.

\section{Thermodynamics of charged dilatonic AdS black holes} \label{sec2}

The Lagrangian of general EMD theory consisting of gravity, a single Maxwell field $A$ and a dilaton field $\phi$ in $D\ge 4$ dimensions is given by
\be
e^{-1} {\cal L}= R-\frac12(\partial\phi)^2 -\frac14 e^{a\phi} F^2 -V(\phi)  \,, \label{lagrangian}
\ee
where $e=\sqrt{-g}$, $F = d A$, $V$ is a scalar potential inspired by gauged supergravity and $a$ is the dilaton coupling constant which can be reparameterized by~\cite{Lu:2013eoa}
\be
a = \sqrt{\frac{4}{N}-\frac{2 (D-3)}{D-2} }   \,. \label{constraint1}
\ee
 The charged dilatonic AdS black holes with $a$ can be viewed as dilatonic AdS black hole with $N$ equal charges~\cite{Duff:1994jr, Rahmfeld:1995fm}. So the value of $N$ should be positive integers as required by supergravity. The reality condition of $a$ requires that
\be
0<N\leq N^{RN} \,, \quad \textup{with} \quad N^{RN}= \frac{2 (D-2)}{D-3} \,.  \label{reality}
\ee
When $N=N^{RN}$, i.e., $a=0$, the dilaton decouples and the theory reduces to EM theory. Although the Lagrangian can be made real by letting $\phi\rightarrow i \phi$ when $N>N^{RN}$, we shall not consider such a situation at all. Due to the maximal dimension allowed by supergravity being seven,  we can easily present all the possible values of $N$ in different dimensions as in Table~\ref{table:Nvalue}.
 \begin{table}[h]
 \centering
\begin{tabular}{|c|c|c|c|c|c|c|}
   \hline  $D$  & 4 & 5 & 6 & 7  \\
   \hline $N$ & 1,2,3,4 & 1,2,3 & 1,2 & 1,2  \\
   \hline
\end{tabular}
   \caption{All possible values of $N$ for gauged supergravities in various dimensions.}   \label{table:Nvalue}
 \end{table}
 The potential can be expressed in terms of a super potential $W$~\cite{Lu:2013eoa},
\be
V= \left( \frac{dW}{d\phi} \right)^2 -\frac{D-1}{2(D-2)}W^2 \,,\quad \textup{with} \quad W= \frac{1}{\sqrt{2}} N (D-3) g \left(e^{-\frac12 a \phi} -\frac{a}{\tilde{a}} e^{-\frac12 \tilde{a} \phi} \right) \,, \label{potential}
\ee
where $ a\  \tilde{a} = -2(D -3)/(D -2)$ and $g$ is the gauge coupling constant~(there should be no confusion between the gauge coupling constant and the determinant of the metric).

The static AdS black hole solutions for Lagrangian~(\ref{lagrangian}) with scalar potential~(\ref{potential}) and constraints~(\ref{constraint1}) are given by~\cite{Lu:2013eoa}
\begin{align}
ds^2 &= -h^{-\frac{D-3}{D-2} N} f dt^2 + h^{\frac{N}{D-2}} \Big(\frac{dr^2}{f}+r^2 d\Omega_{D-2}^2\Big) \,,\\
A &=\sqrt{\frac{N(m+q)}{q}} h^{-1} dt \,,\quad \phi = \frac12 N a \log h \,,\\
f &= 1-\frac{m}{r^{D-3}}+ g^2 r^2 h^{N} \,, \quad h = 1+\frac{q}{r^{D-3}} \,,
\end{align}
where parameters $m$ and $q$ characterize mass and electric charge, and $d\Omega_{D-2}^2$ represents the unit $(D-2)$-sphere, the $(D-2)$-torus or the hyperbolic $(D-2)$-space. The topological black holes can easily be obtained by some appropriate scaling. In our work, we only consider spherical black holes for simplicity.

The event horizon of black hole is determined by the largest (real) root of $f(r_0) = 0$. The thermodynamic quantities are given by~\cite{Lu:2013eoa}
\begin{align}
M &= \frac{\pi^{\frac{D-3}{2}}}{8 \, \Gamma[\frac{D-1}{2}]} \left( (D-2) m + (D-3) N q \right) \,,\\
T &= \frac{f'}{4 \pi  h^{\frac{N}{2}}}  \,,  \qquad S = \frac{1}{4} {\cal A} = \frac{\pi^{\frac{D-1}{2}} r_0^{D-2}}{2 \, \Gamma[\frac{D-1}{2}]} h^{\frac{N}{2}} \,,  \label{temp} \\
Q &= \frac{(D-3) \pi^{\frac{D-3}{2}}}{8 \, \Gamma[\frac{D-1}{2}]} \sqrt{N q (m+q)} \,, \qquad \Phi = \sqrt{\frac{N (m+q)}{q}} \left(1-\frac{1}{h} \right) \,, \label{charge}
\end{align}
where $\Gamma$ indicates the Gamma function, and $(M, T, S, Q, \Phi)$ denote mass, temperature, Bekenstein-Hawking entropy, electric charge and electric potential, respectively. Further, the gauge coupling constant (cosmological constant) can be interpreted as the thermodynamical pressure ${\cal P}$ in the extended phase space,
\be
{\cal P} = -\frac{1}{8 \pi} \Lambda = \frac{(D-1)(D-2)}{ 16 \pi \ell^2}  = \frac{(D-1)(D-2)}{ 16 \pi} g^2  \,. \label{pressure}
\ee
The corresponding thermodynamic volume is given by
\be
\begin{split}
{\cal V}
= \frac{2 \pi^{\frac{D-1}{2}} \left(q + r_0^{D-3}\right)^{N-1} }{(D-1)\, \Gamma(\frac{D-1}{2})} \left(  r_0^{D-3} + \Big(1- \frac{N}{N^{RN}} \Big) q \right) r_0^{(D-1)-(D-3)N} \,.
\end{split} \label{vol1}
\ee
The thermodynamic volume satisfies the ``Reverse Isoperimetric Inequality'' conjecture~\cite{Cvetic:2010jb}. Although the conjecture is not proven, it has been checked in most black holes and follows from null-energy condition~\cite{Feng:2017wvc}.
These quantities satisfy the following first law and Smarr relation
\bea
d M &=& T dS +\Phi dQ +{\cal V} d{\cal P}  \,,\\
M &=& \frac{D-2}{D-3} T S + \Phi Q + \frac{2}{D-3} {\cal V} d{\cal P} \,.
\eea
For our purposes, to study the ${\cal P - V}$ phase transitions of the charged dilatonic AdS black holes in the canonical ensemble of extended phase space, we fix the electric charge $Q$ and treat it not as a thermodynamical variable in the subsequent discussion. So the mass $M$ can be viewed as the enthalpy $H$,
\be
d M \equiv d H = T dS +{\cal V} d{\cal P} \,.
\ee
In order to study the ${\cal P - V}$ phase transitions, it is convenient to introduce a special volume $v$ by analyzing the dimensional scaling~\cite{Kubiznak:2012wp}. By assuming the shape of black hole to be a regular sphere, the special volume~\cite{Gunasekaran:2012dq} can be viewed as the effective radius of the black hole,
\be
\frac{2 \pi^{(D-1)/2}}{(D-1) \Gamma(\frac{D-1}{2})} v^{D-1 } = {\cal V} \quad  \Rightarrow \quad   v = \Big(\frac{(D-1) \Gamma(\frac{D-1}{2}) {\cal V}}{2 \pi^{(D-1)/2} } \Big)^{\frac{1}{D-1}}  \,.
\ee
So we can study the ${\cal P}-v$ criticality instead of ${\cal P-V}$ criticality in the subsequent discussions.

\section{${\cal P}-v$ criticality }  \label{sec3}

The ${\cal P}-v$ phase transition  behaviors of four dimensional charged dilatonic AdS black holes with $N=1,2,3,4$, which can be viewed as STU black holes with single charge, two equal charges, three equal charges and four equal charges (or RN AdS black hole), have been studied in~\cite{Caceres:2015vsa}. The behaviors of the four cases are different from each other. So we will explore whether the transition behaviors of higher dimensional dilatonic AdS black holes with $N$ equal charges show the same behaviors as their four dimensional counterparts. The behavior of a five dimensional charged dilatonic AdS black hole with $N=3$, which can be viewed as a RN AdS black hole, has also been studied in\cite{Gunasekaran:2012dq}. The dilaton decouples and the theory reduces to EM theory. So we focus on the cases $N=1$ and $N=2$  in five, six and seven dimensions. We find that the transition behaviors can also be changed by dimension. There emerge new behaviors in higher dimensions which are absent in four dimensions.

\subsection{$N=2$ cases}   \label{sec3.1}

\begin{figure}[h]
\centerline{
\ \ \ \ \ \includegraphics[width=0.45\linewidth]{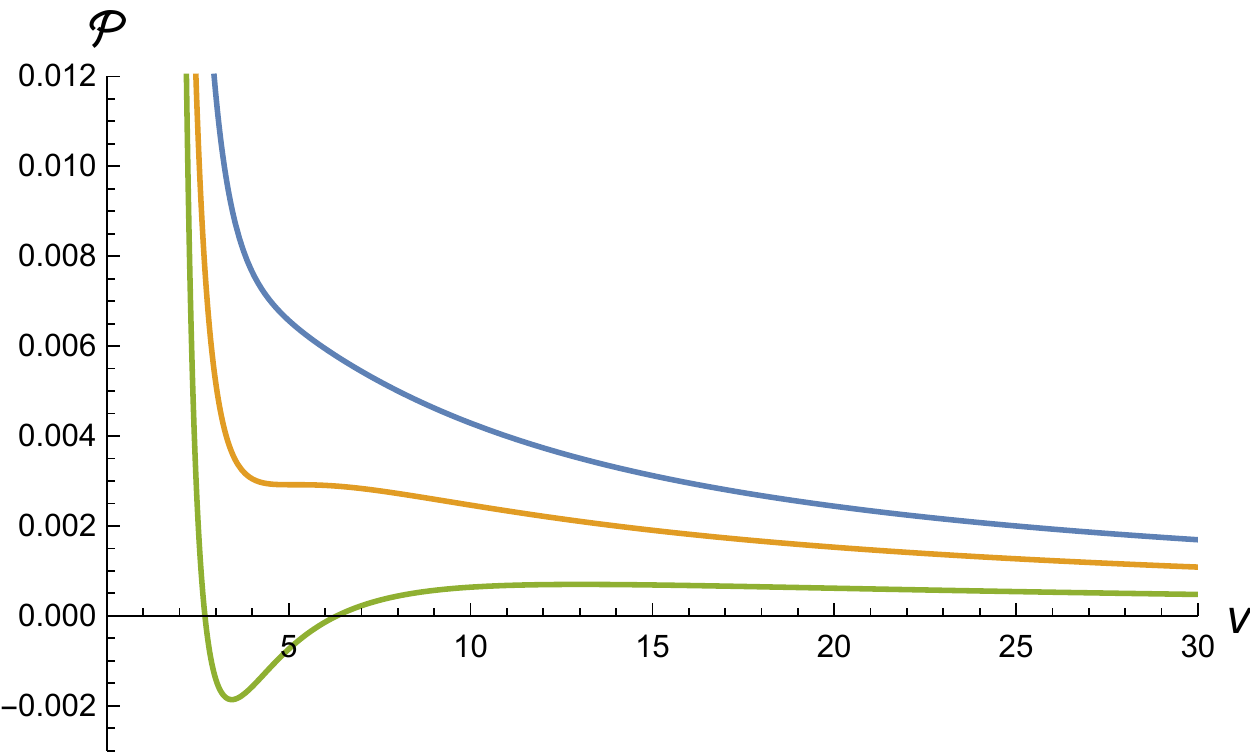}\ \ \ \
\includegraphics[width=0.45\linewidth]{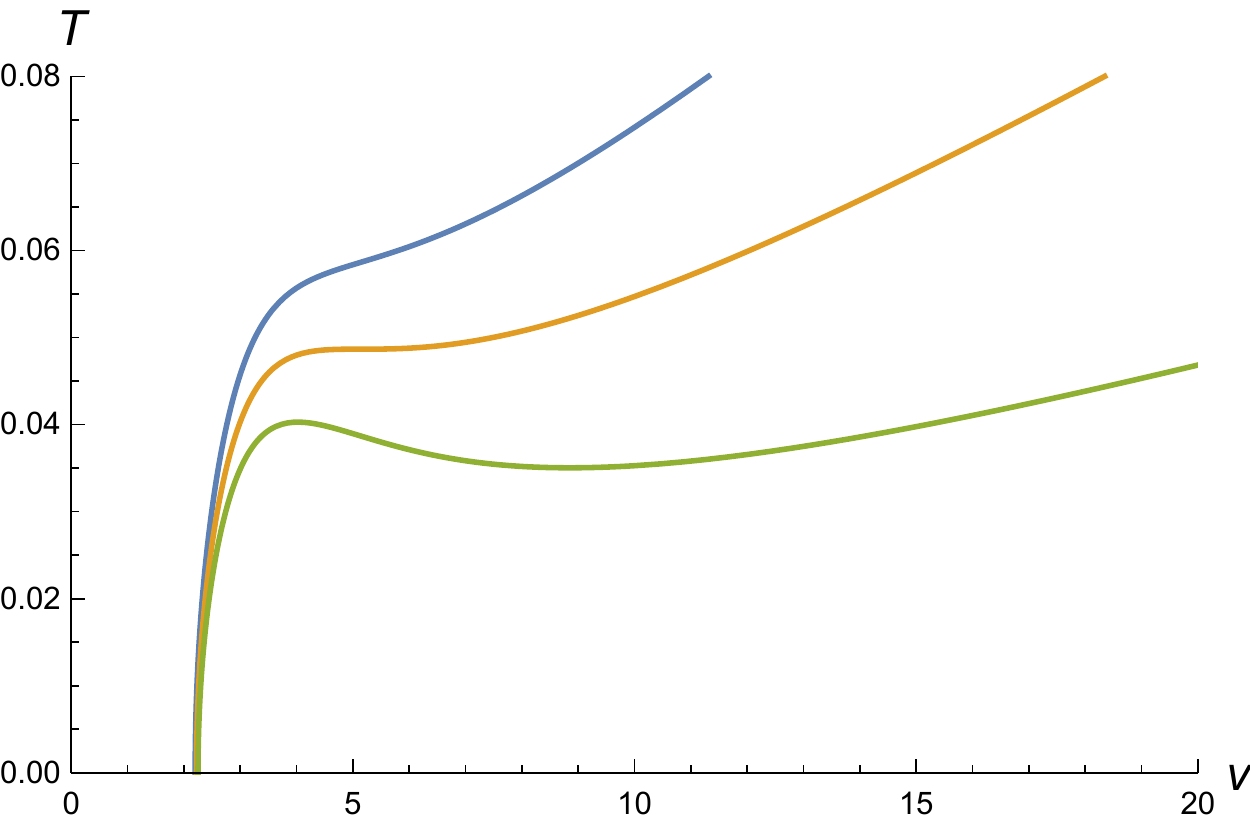}}
\caption{${\cal P } - v$ diagram (left) and $T - v$ diagram (right) of charged dilatonic AdS black hole with $N=2, D=5$.   We set electric charge $Q=10$. In ${\cal P } - v$ diagram, the lines represent isotherms with temperatures $3 T_c/2, T_c, T_c/2$ from top to bottom. In $T - v$ diagram, the lines represent isobars with pressures $3 {\cal P }_c/2, {\cal P }_c, {\cal P }_c/2$ from top to bottom.}
\label{pv52}
\end{figure}

\noindent{\bf $D=5$ case}: As shown in Ref.~\cite{Caceres:2015vsa}, the thermodynamic pressure goes to zero as the thermodynamic volume decreases to zero at sufficiently low temperature for the four dimensional charged dilatonic AdS black holes with $N = 2$. As the temperature increases, there is a phase transition at some critical temperature and the thermodynamic pressure goes to infinity as the thermodynamic volume decreases to zero in the system at sufficiently high temperature. The behavior in five dimensions is rather different from the four dimensional counterpart. From the analysis of the previous section it is hard to obtain the analytic expression of equation of state because of the coupling of different thermodynamic quantities. However, we can analyze the phase transitions by numeric method. We plot the ${\cal P } - v$ and  $T - v$ diagrams in Fig.~\ref{pv52}.
 The corresponding critical point can be obtained by
 \be
 \frac{\partial {\cal P}}{\partial v} \Big|_{T_c} = 0 \,, \quad  \frac{\partial^2 {\cal P}}{\partial v^2} \Big|_{T_c}  = 0 \,, \quad \textup{or}  \quad \frac{\partial T}{\partial v} \Big|_{{\cal P}_c} = 0 \,, \quad  \frac{\partial^2 T}{\partial v^2} \Big|_{{\cal P}_c}  = 0 \,. \label{cp}
 \ee
 The critical point and the universal ratio independent of $Q$ are given by
 \be
 T_c = 0.154\ Q^{-\frac12}\,, \quad v_c = 1.63\ Q^{-1}\,, \quad {\cal P}_c = 0.0292\ Q^{\frac12}\,, \quad \frac{{\cal P }_c v_c}{T_c} = 0.308 \,.
 \ee
 From Fig.~\ref{pv52}, we find the ${\cal P}-v$  transitions of five dimensional charged dilatonic AdS black holes with $N = 2$ have standard Van der Waals behavior which is different from the four dimensional dilatonic AdS black hole with two equal charges~\cite{Caceres:2015vsa}. Actually, the four dimensional charged dilatonic AdS black holes with $N = 3$ also have similar behaviors~\cite{Caceres:2015vsa}. However, we study the ${\cal P}-v$ transitions for all possible values of $N$ supported by gauged supergravities, and the same behaviors do not appear again. For more details as regards the standard Van der Waals transitions of these cases, we refer to our future work~\cite{Li:2018aax}.

\begin{figure}[h]
\centerline{
\ \ \ \ \ \includegraphics[width=0.9\linewidth]{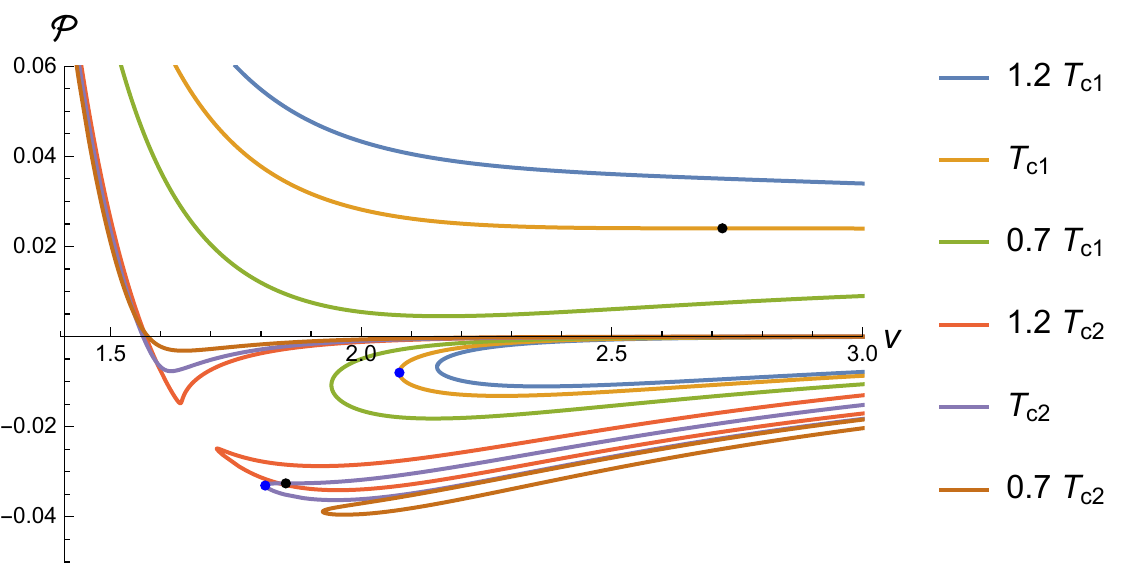}}
\caption{Isotherms in ${\cal P} - v$ diagram of charged dilatonic AdS black hole with $N=2, D=6$.  We set electric charge $Q=10$. The two black points represent the critical points and the two blue points represent incompressible points at critical temperatures.}
\label{pv62}
\end{figure}

\noindent{\bf $D=6$ and $D=7$ cases}: Now we study the ${\cal P}-v$  transitions in six and seven dimensions. Unlike the counterparts in $ D=4$ and $ D=5$, the charged dilatonic AdS black holes in six and seven dimensions have new Van der Waals-like behaviors. By solving Eq.~(\ref{cp}), we obtain two critical points, which are given in Table~\ref{table:tprhovalue}.
 \begin{table}[h]
 \centering
$\begin{array}{|c|c|c|c|c|c|c|}
   \hline  D  & T_{c1} & T_{c2} & {\cal P }_{c1} & {\cal P }_{c2} & v_{c1} & v_{c2}   \\
   \hline \quad 6 \quad & 0.323\ Q^{-1/3} & 0.0697\ Q^{-1/3} & 0.112\ Q^{-2/3} & -0.151\ Q^{-2/3} & 1.26\ Q^{1/3} & 0.859\ Q^{1/3}  \\
   \hline 7 & 0.495\ Q^{-1/4}  & 0.109\ Q^{-1/4} & 0.243\ Q^{-1/2} & -0.279\ Q^{-1/2} & 1.14\ Q^{1/4} & 0.847\ Q^{1/4}   \\
   \hline
\end{array}$
   \caption{Two critical points with $N=2$ in six and seven dimensions.}   \label{table:tprhovalue}
 \end{table}
 It is interesting that one of the critical pressure is negative and the behavior was not obtained in previous related research to the best of our knowledge. Besides the negative pressure, all the physical quantities of black holes are positive, such as mass, entropy, temperature and so on.  The exponent of $Q$ of the expression of the temperature, the thermodynamic pressure and the volume at the two critical points are the same. So the negative pressure could be explained analogously as the repulsive force. Actually, similar interesting phase transition phenomena at negative pressure have been observed in the liquid-liquid system, for example,~\cite{Vasisht}. The ratios are independent of $Q$ in both cases which are given by
 \bea
 D&=&6: \quad \frac{{\cal P }_{c1} v_{c1}}{T_{c1}} = 0.436\,,\quad  \frac{{\cal P }_{c2} v_{c2}}{T_{c2}} = -1.86 \,,\\
 D&=&7: \quad \frac{{\cal P }_{c1} v_{c1}}{T_{c1}} = 0.562\,,\quad  \frac{{\cal P }_{c2} v_{c2}}{T_{c2}} = -2.17 \,.
 \eea
Because the Van der Waals-like behaviors of six and seven dimensional dilatonic AdS black holes with two equal charges are the same  (a similar behavior  exists in $D>7$ dimensions, which are not supported by gauged supergravity theory, so we do not consider the higher dimensional cases in this paper), so we only plot the ${\cal P } - v$ diagram with $N=2$ of the six dimensional charged dilatonic AdS black hole which is given by Fig.~\ref{pv62}.
 There are two branches of the isotherms in the ${\cal P } - v$ diagram at a certain temperature. At sufficiently high temperature, the upper branch of the isotherms undergoes a normal Van der Waals phase transition, which also appears in RN AdS black holes. As the temperature decreases, two branches intersect and exchange part of their lines. Then at sufficiently low temperature, the lower branch of isotherms also shows Van der Waals-like behavior. We should mention that the lower branch of the isotherm will disappear at sufficiently low temperature. The thermodynamic pressure goes to positive infinity as the thermodynamic volume goes to zero. The thermodynamic pressure goes to zero as the thermodynamic volume goes to infinity.

\begin{figure}[h]
\centerline{
\ \ \ \ \ \includegraphics[width=0.75\linewidth]{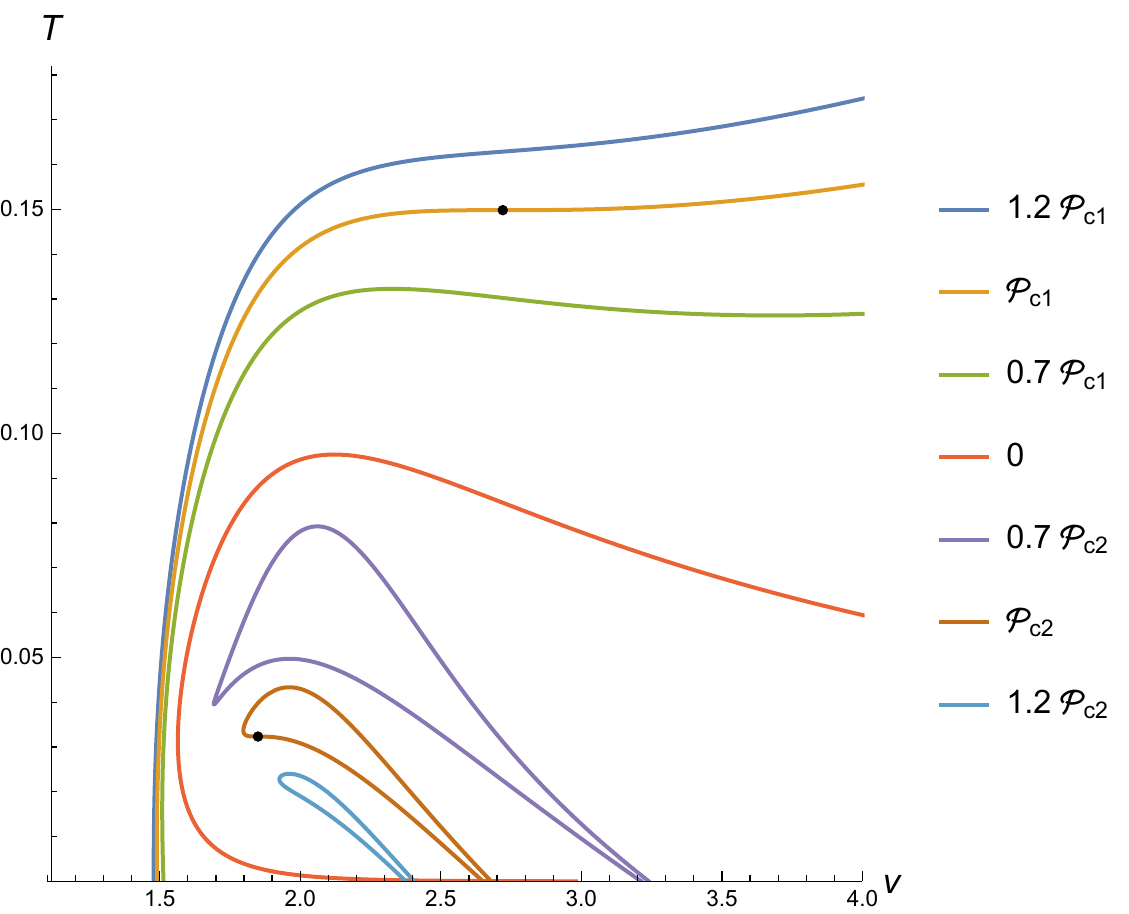}}
\caption{ $T - v$ diagram of dilatonic black hole with $N=2, D=6$. We set electric charge $Q=10$.  The two black points represent the critical points.}
\label{tv62}
\end{figure}

\begin{figure}[h]
\centerline{
\ \ \ \ \ \includegraphics[width=0.9\linewidth]{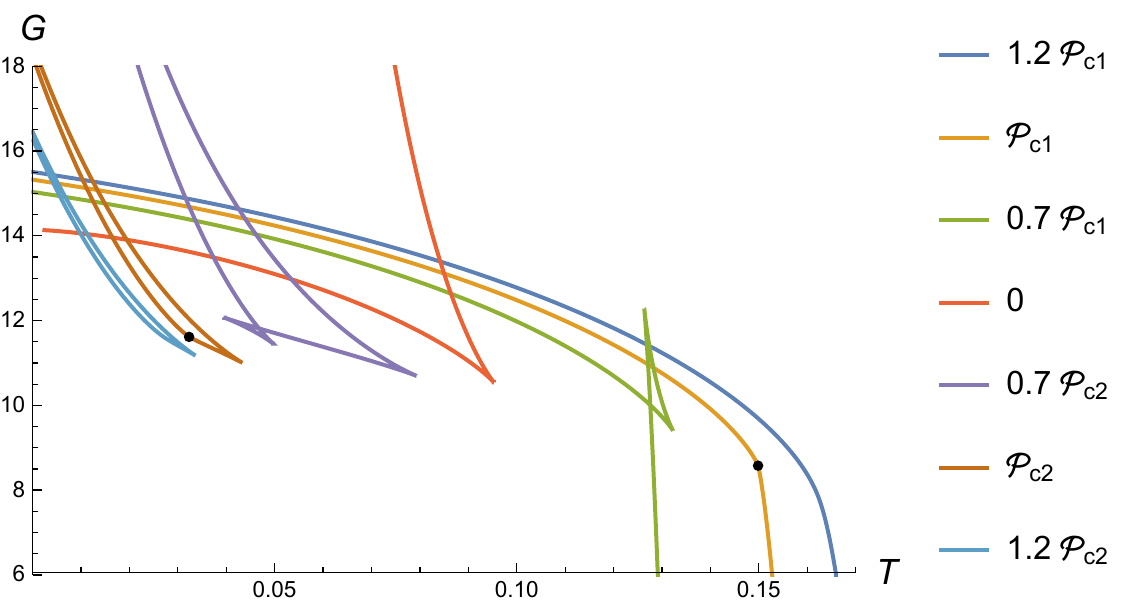}}
\caption{ $G - T$ diagram of charged dilatonic AdS black hole with $N=2, D=6$.   We set electric charge $Q=10$. The black points represent the critical points.}
\label{gt62}
\end{figure}

\begin{figure}[h]
\centerline{
\ \ \ \ \ \includegraphics[width=0.5\linewidth]{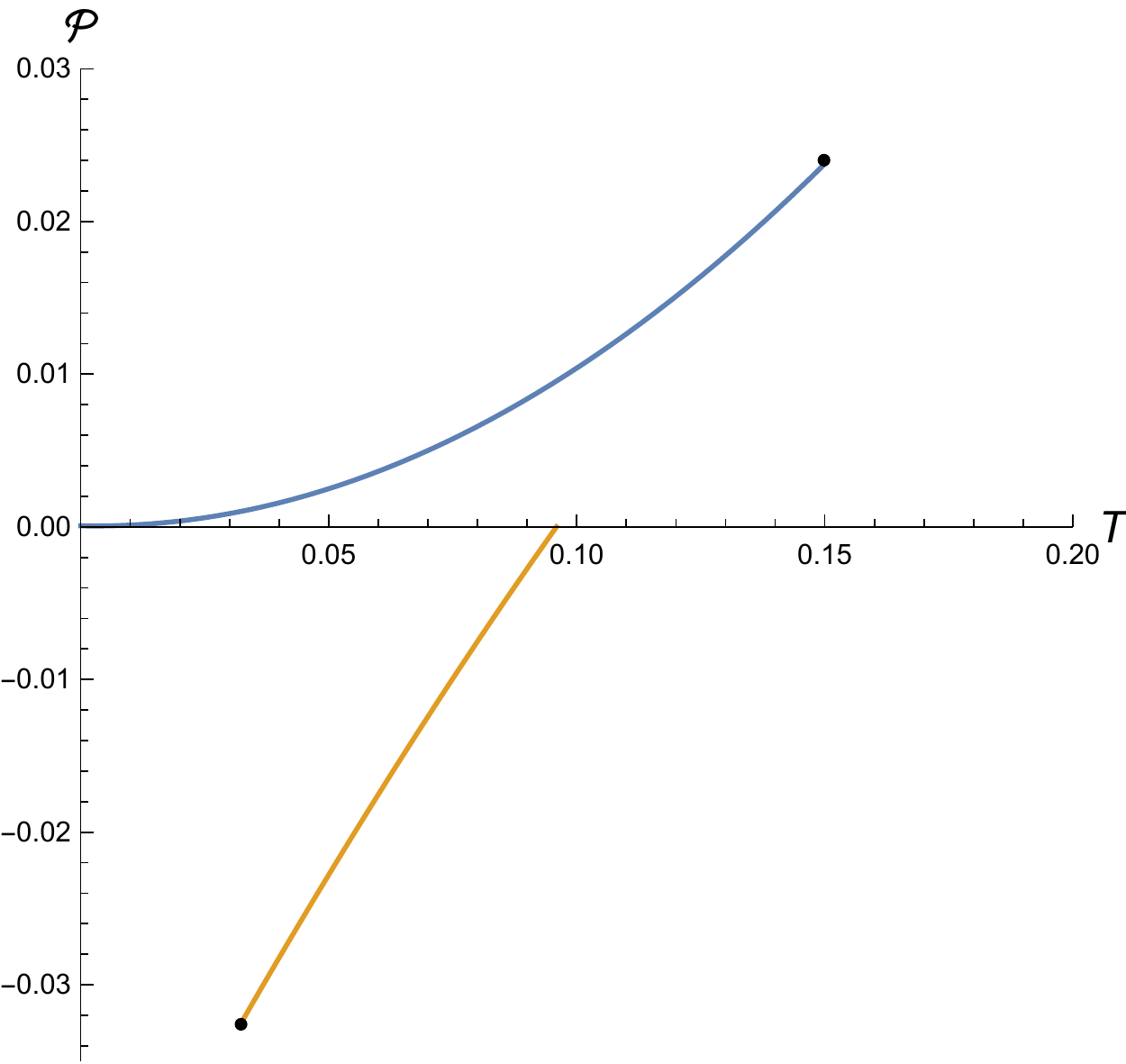}}
\caption{${\cal P} - T$ plane of dilatonic black hole with $N=2, D=6$. We set the electric charge $Q=10$. In the plane, the top and bottom lines represent the positive and negative pressure zones. The black points represent the critical points.}
\label{pt62}
\end{figure}

 We find that there exists another interesting feature, the incompressible points, in the lower branch of isotherms. The incompressible point can be obtained by~\cite{Dolan:2011jm}
 \be
 -\frac{1}{v} \frac{\partial {\cal P }}{\partial v} = 0  \,,
 \ee
and the critical pressures and special volumes of two incompressible points at critical temperatures are given by
\be
{\cal P }_{\textup{inc}1} =-0.0112 \,,\quad v_{\textup{inc}1} =1.69 \,,\quad \textup{and} \quad  {\cal P }_{\textup{inc}2} =-0.174 \,,\quad v_{\textup{inc}2} =1.34 \,.
\ee
We also plot the isobars of $T - v$ diagram in Fig.~\ref{tv62} which provides another perspective on the study of the Van der Waals-like phase transition.
 The Gibbs free energy can be obtained simply by $G \equiv G(T, {\cal P })= M - T S$. According to~\cite{Kubiznak:2012wp}, we also plot the isobars of $G - T$ diagram in Fig.~\ref{gt62}.
From Fig.~\ref{gt62}, we  find that there also exists a swallowtail behavior at negative pressure which is a typical feature characterizing the multivalued function~(it may appear in many other systems, e.g.~\cite{Feng:2018qnx}). The swallowtails disappear at both positive and negative critical pressures. The critical exponentials for the critical point with positive critical pressure are given by
\be
\alpha =0 \,,\quad \beta = \frac{1}{2} \,,\quad \gamma =1 \,,\quad \delta = 3 \,,
\ee
which are same as the case of the standard Van der Waals behavior. So we can definitively claim that the behavior under the extended phase space assumption with positive critical pressure is of Van der Waals type. We also calculate the critical exponentials at the negative critical pressure. The values of $\alpha, \beta$ and $\delta $ are the same.  However, the isothermal compressibility obeys $\kappa_T = -\frac{1}{v} \frac{\partial v}{\partial {\cal P}}|_T <0$ which cannot be used to obtain  $\gamma $. On taking the absolute value of $\kappa_T$, the value of $\gamma$ is given by 3.  We plot the ${\cal P} - T$ plane in Fig.~\ref{pt62}.
The Van der Waals-like behavior in the negative pressure can also be viewed in terms of small/large black hole transitions.

\begin{figure}[h]
\centerline{
\ \ \ \ \ \includegraphics[width=0.45\linewidth]{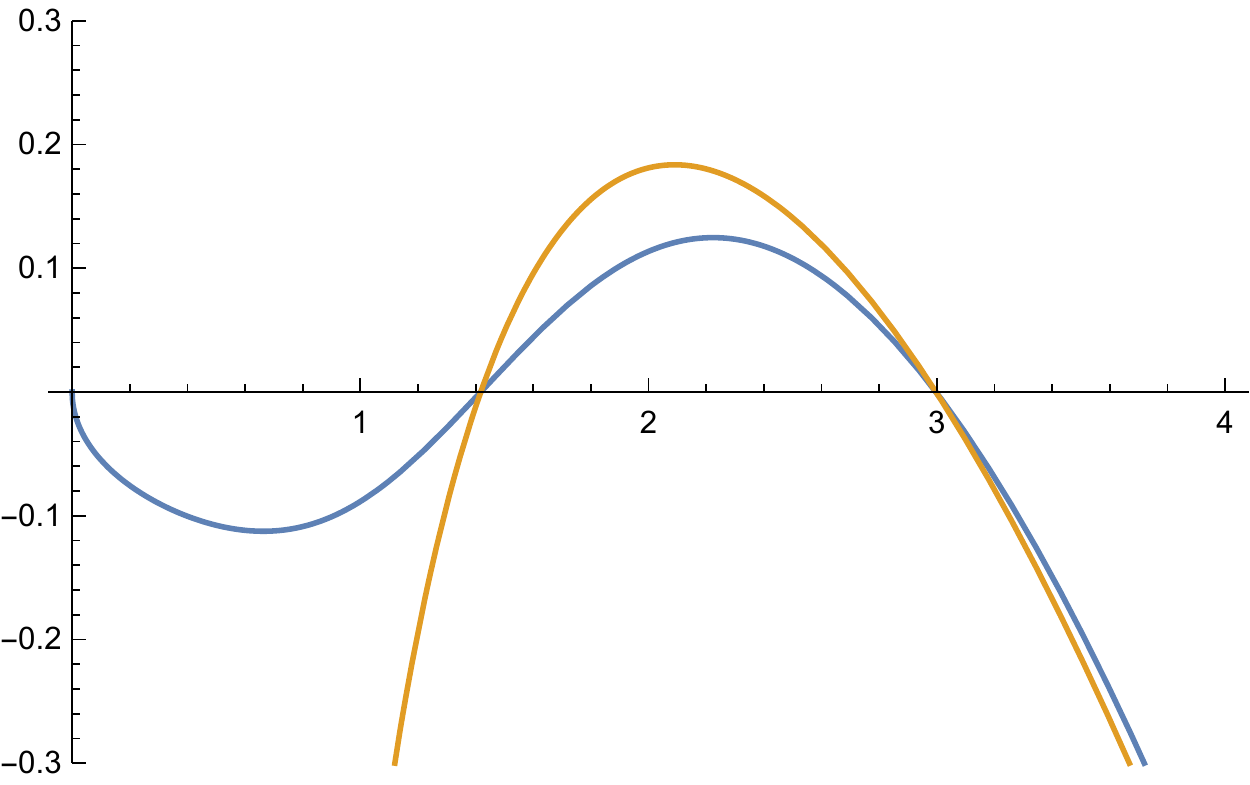}}
\caption{ The horizontal axis is $r_0$, the blue line and orange line are $ -g_{tt} $ and  $ g_{rr}^{-1}$  of the charged dilatonic black hole for $N=2$ in six dimensions at negative critical point respectively.}
\label{horizon62}
\end{figure}

 Here we present some discussion of about the new transition behavior mentioned above. First of all, the high dimensional dilatonic AdS black holes with the new transition exist indeed. During the numerical analysis, the radius of the  horizon $r_0$ should be the outer horizon rather than the inner horizon.  In the negative pressure region, AdS black holes convert to de Sitter (dS) black holes. For dS black holes, there exists another cosmological horizon larger than the event horizon. From Fig.~\ref{horizon62} ,
the first intersection of $-g_{tt}$ and $g^{-1}_{rr}$ is the outer horizon and the second one is the cosmological horizon. So we can say the black holes exist indeed. Actually, the transition behavior of the critical point is similar to the phase transition in the RN-dS black hole which is briefly given in Appendix~\ref{app}. So the phase transition may not be connected with AdS/CFT correspondence. The boundary field interpretation should be studied further. Second, all the analyses of extended thermodynamics are based on the assumption that the cosmological constant can be viewed as thermodynamic pressure and the corresponding conjugate quantity is the thermodynamic volume. In the mathematical sense, the existence of the new transition relates to the idea that the expressions of the thermodynamic volume of dilatonic AdS black holes do not entail the regular volume definition in the sense of the radius of the horizon $r_0$. And the real definition of the volume of the black hole is still unclear. So a thorough explanation of the new transition behavior might rely on the further study of the pressure and volume of black holes. Third, it should be noted that the four dimensional EMD theory can be obtained by performing dimensional reduction\cite{Lu:1998xt,Cremmer:1999du} and truncation on the higher dimensional supergravities. The new behavior does not exist in the four dimensional dilatonic AdS black holes with any equal charges. So it is worth to study the new transition behavior further via dimensional reduction.

\subsection{$N=1$ cases}   \label{sec3.2}

\begin{figure}[h]
\centerline{
\ \ \ \ \ \includegraphics[width=0.45\linewidth]{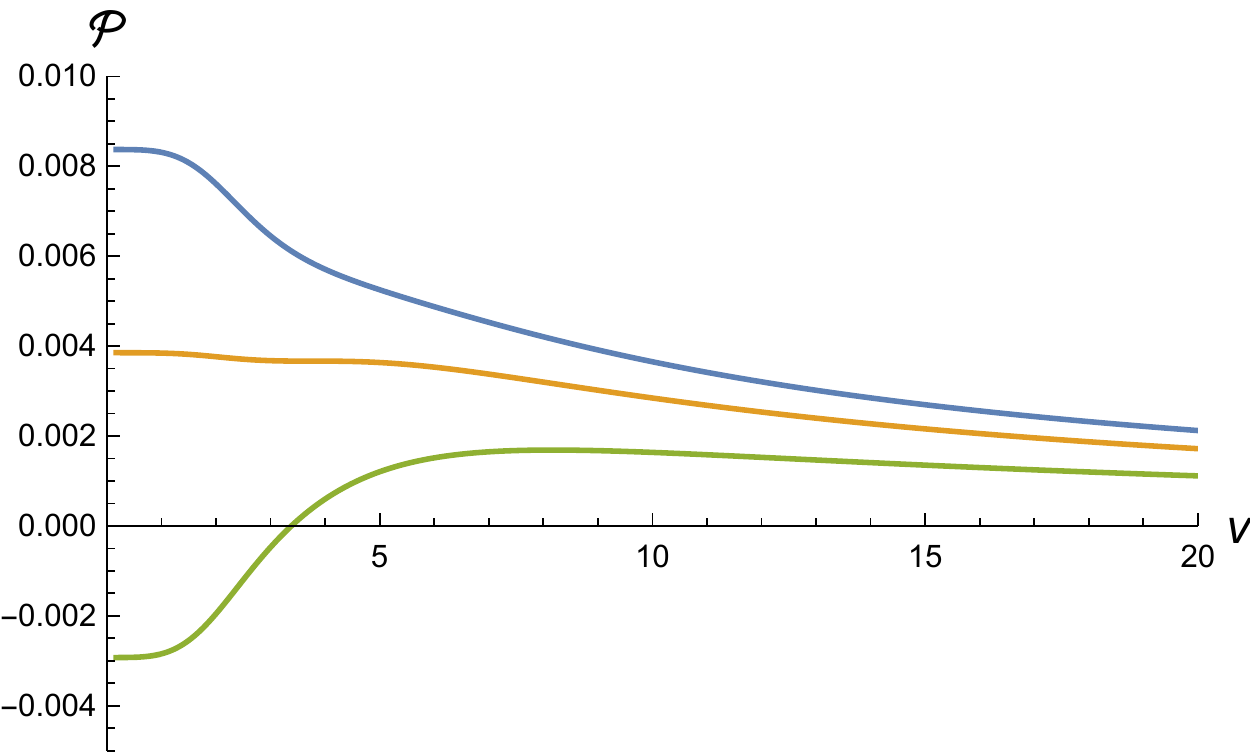}\ \ \ \
\includegraphics[width=0.45\linewidth]{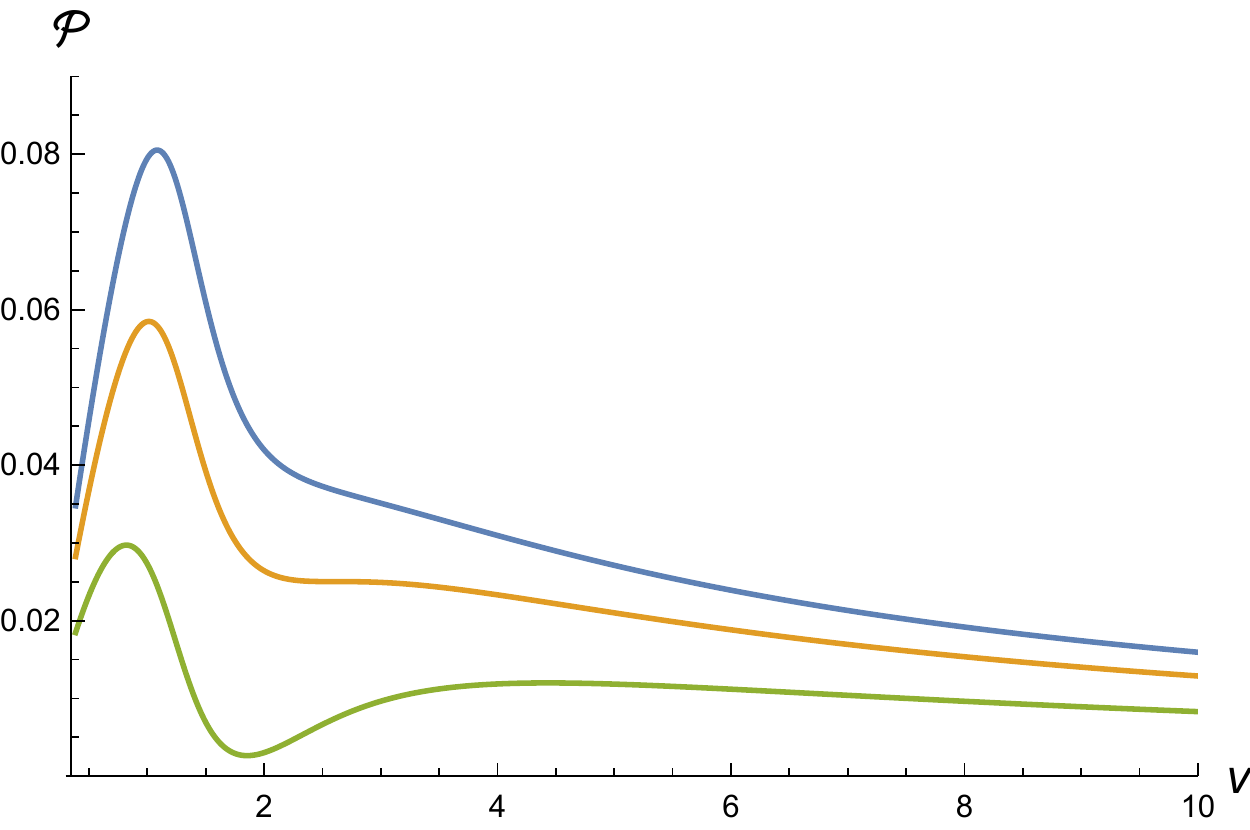}}
\caption{${\cal P} - v$ diagrams of charged dilatonic AdS black hole for $N=1$ in five and six dimensions.  We set electric charge $Q=10$. The temperatures of the lines from top to bottom are $T= 1.2\ T_{c}\,, T_{c}$, and  $0.7\ T_{c} $ respectively.}
\label{pv51}
\end{figure}

Now we study the ${\cal P} - v $ phase transitions of charged dilatonic AdS black holes with $N=1$ in five, six and seven dimensions. The $N=1$ supergravities can be viewed as Kaluza-Klein theory. As shown in Ref.~\cite{Caceres:2015vsa}, there is no Van der Waals transition in four dimensions. The thermodynamic pressure goes to negative infinity as the volume decreases to zero which is the same as the Hawking-Page transition. However, the behavior is different in higher dimensions. We can obtain the critical points from Eq.~(\ref{cp}) and the universal ratios are given by Table~\ref{table:tprhon1}.
 \begin{table}[h]
 \centering
$\begin{array}{|c|c|c|c|c|}
   \hline  D  & T_{c} & {\cal P }_{c} & v_{c} &  \frac{{\cal P }_{c} v_{c}}{T_{c}}  \\
   \hline \quad 5 \quad & \quad 0.170\ Q^{-1/2} \quad & \quad 0.0367\ Q^{-1} \quad & \quad 1.19\ Q^{1/2} \quad & \quad 0.256 \quad   \\
   \hline \quad 6 \quad & 0.329\ Q^{-1/3} & 0.116\ Q^{-2/3} & 1.21\ Q^{1/3} & 0.428  \\
   \hline \quad 7 \quad & 0.499\ Q^{-1/4} & 0.247\ Q^{-1/2} & 1.13\ Q^{1/4} & 0.558   \\
   \hline
\end{array}$
   \caption{Critical points with $N=1$ in five, six and seven dimensions.}   \label{table:tprhon1}
 \end{table}
 We plot the ${\cal P} - v $ diagrams of $N=1$ in five and six dimensions in Fig.~\ref{pv51}~(the behavior in $D>6$ is similar to the six dimensional counterpart, so we do not plot them here).
We find the thermodynamic pressure does not go to positive infinity as the thermodynamic volume decreases to zero. However, there exist Van der Waals-like transition behaviors in $D>4$ dimensions which are absent in four dimension. The $G - T$ diagrams are similar to the ordinary RN case, so we do not plot them here.

\section{Conclusions and outlook} \label{sec4}
By studying the extended thermodynamics of charged dilatonic AdS black holes in a class of EMD theory inspired by gauged supergravities in various dimensions, we find that the transition behaviors of higher dimensional black holes with $N$ equal charges are different from their four dimensional counterparts. There emerge some new interesting Van der Waals (-like) transition behaviors in higher dimensions. Firstly, we find that there are two cases $D=4, N=3$ and $D=5, N=2$ with standard Van der Waals transition behaviors in gauged supergravities. Second, we find there are two branches of the transition in $D>5$ dimensions with $N=2$. One is similar to Van der Waals behavior, and another one is new behavior in which the critical pressure is negative. In the new branch transition, there exists an incompressible point. Though the isothermal compressibility of the critical point is not compatible with ordinary thermodynamics, it is still significant for the study of the new phase transitions in both boundary and bulk view of gauge / gravity duality. Third, we find there exist Van der Waals -like transitions in  $N=1$ case in $D>4$ dimensions which are absent in four dimensions.

Due to the rich phase behaviors of charged dilatonic AdS black holes, it is valuable to study the extended thermodynamics of more general AdS black holes, such as rotating~\cite{Wu:2011zzh} or dyonic~\cite{Lu:2013ura,Chow:2013gba} black holes in ($\omega$-deformed) gauged supergravity systems~\cite{Lu:2014fpa,Wu:2015ska}.

\section*{Acknowledgement }

SLL is grateful to Shuang-Qing Wu and Zhuo Cai for kind discussions. He thanks especially H. L\"u for many suggestions and proofreading. SLL, HKD and HW are  supported in part by NSFC grants No.~11575022, No.~11175016 and Graduate Technological Innovation Project of Beijing Institute of Technology. HDL is supported in part by NSFC grants No.~11875200 and No.~11475024.

\appendix

\section{${\cal P } - v$ transition of RN-dS black hole} \label{app}

\begin{figure}[h]
\centerline{
\ \ \ \ \ \includegraphics[width=0.45\linewidth]{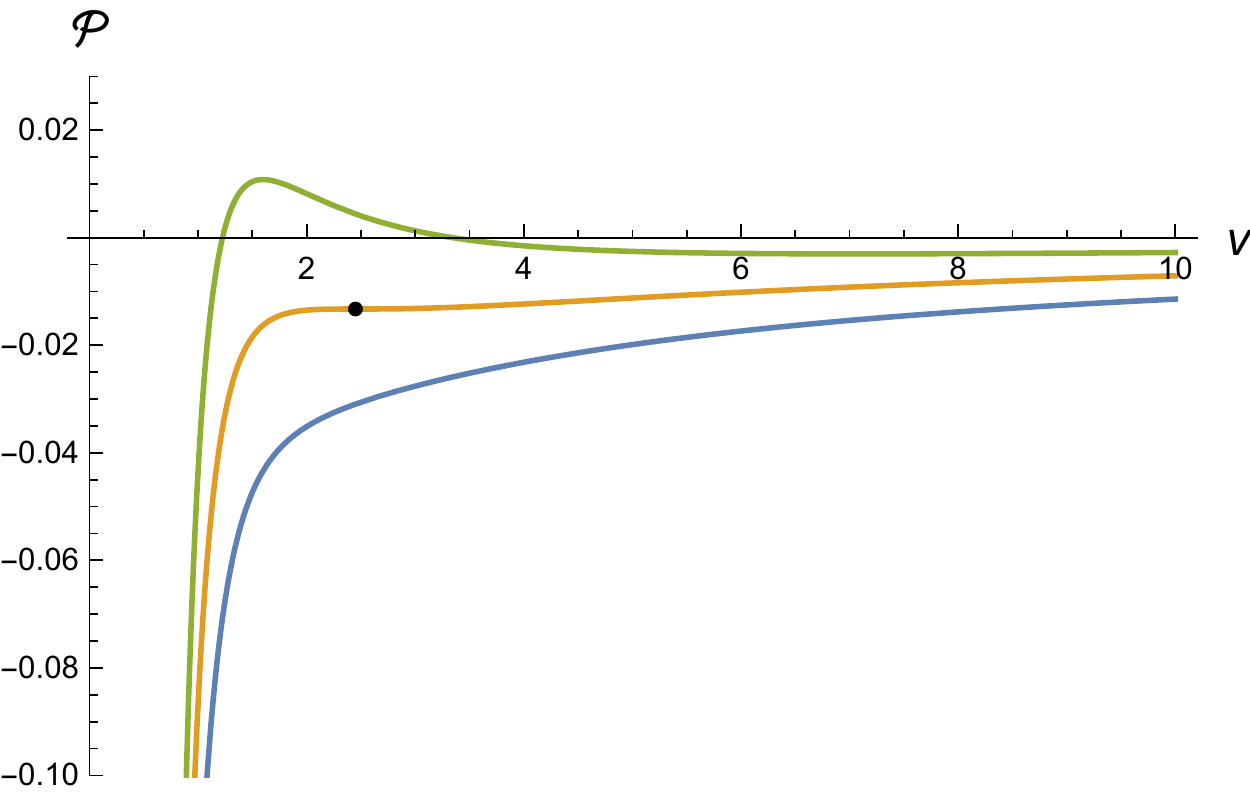}}
\caption{${\cal P} - v$ diagram of the RN-dS black hole in $D=4$ dimensions.  We set $q=1$. The temperatures of the lines from top to bottom are $T= 1.5\ T_{c}\,, T_{c}$, and  $0.5\ T_{c} $ respectively. The black point is the critical point.}
\label{pvds}
\end{figure}

The RN-dS black hole is given by
\be
ds^2 =  -(1- \frac{2 m}{r} -\frac{r^2}{\ell^2} +\frac{q^2}{4 r^2}) dt^2 +(1- \frac{2 m}{r} -\frac{r^2}{\ell^2} +\frac{q^2}{4 r^2})^{-1} dr^2 +r^2 d\Omega_2^2 \,.
\ee
The exact EoS is given by
\be
{\cal P} = -\frac{T}{v} +\frac{1}{2 \pi v^2} -\frac{q^2}{2 \pi v^4}  \,,
\ee
where $v=r_0/2$~\cite{Kubiznak:2012wp}.
The ${\cal P}-v$ diagram is given by Fig.~\ref{pvds}.

\end{document}